\documentclass[pdflatex,sn-vancouver-num]{sn-jnl}
\usepackage[T1]{fontenc}
\usepackage[utf8]{inputenc}
\usepackage{lmodern,ragged2e}
\usepackage{graphicx,amsmath,amssymb,booktabs,array,longtable,xcolor}
\usepackage{xurl}
\usepackage[compatibility=false]{caption}
\makeatletter
\renewcommand\normalsize{\@setfontsize\normalsize{12}{14.5}\abovedisplayskip 12pt plus2pt minus1pt\belowdisplayskip\abovedisplayskip\abovedisplayshortskip 3pt\belowdisplayshortskip 6pt\let\@listi\@listI}
\renewcommand\small{\normalsize}
\renewcommand\footnotesize{\normalsize}
\def\addressfont{\normalfont\normalsize\titraggedcenter}
\def\abstractheadfont{\normalfont\normalsize\bfseries\titraggedcenter}
\def\abstractfont{\normalfont\normalsize\leftskip=0pt\rightskip=0pt\parfillskip=0pt plus1fil}
\def\keywordfont{\normalfont\normalsize\leftskip=0pt\rightskip=0pt plus0.5fill}
\def\Artcatfont{\normalfont\normalsize}
\makeatother
\normalsize
\newcolumntype{P}[1]{>{\RaggedRight\arraybackslash}p{#1}}
\setcitestyle{numbers,square,comma}
\hypersetup{hidelinks,pdftitle={Reconstructing False Alarm and Detection Rates in Quantum Machine Learning for Intrusion Detection}}
\newcommand{\doi}[1]{\url{https://doi.org/#1}}
\newcommand{\FPR}{\mathrm{FPR}}\newcommand{\TPR}{\mathrm{TPR}}
\newcommand{\FP}{\mathrm{FP}}\newcommand{\FN}{\mathrm{FN}}
\newcommand{\TN}{\mathrm{TN}}\newcommand{\TP}{\mathrm{TP}}
\begin{document}
\articletype{Research}
\title{Reconstructing False Alarm and Detection Rates in Quantum Machine Learning for Intrusion Detection}
\author*[1]{\fnm{Roberto} \sur{Fern\'andez-Barrios}}\email{roberto.fernandez.b@deusto.es}
\author[1]{\fnm{Iker} \sur{Pastor-L\'opez}}
\author[1]{\fnm{Asier} \sur{Gonz\'alez-Santocildes}}
\author[1]{\fnm{Pablo} \sur{Garc\'ia Bringas}}
\affil[1]{\orgdiv{Faculty of Engineering}, \orgname{University of Deusto}, \orgaddress{\street{Avda. de las Universidades, 24}, \city{Bilbao}, \postcode{48007}, \country{Spain}}}
\abstract{Published results in quantum machine learning for intrusion detection (QML-IDS) require joint interpretation of false-positive rate (FPR), attack true-positive rate (TPR), and evaluation provenance. We reconstruct operating points in the 25-entry frame selected by Kaissar and colleagues: 21 included articles, 20 identified documentary families, substantive methodological material consulted for 15, and 24 descriptive points from five. In one test-labelled comparison, a quantum support vector machine reduces FPR from 6.875\% for random forest to 3.329\%, while TPR decreases from 93.955\% to 93.682\%; both rates improve against three other comparators. In another study, a complete benign row identifies FPR as 12/6000 (0.200\%), despite incomplete attack-side allocation that requires additional assumptions for TPR. Normalized matrices identify detection ranges rather than a unique aggregate rate. A uniform provenance rule separates training-described and test-adjusted results from six conditional illustrations whose independence remains unverified. The contribution is a reproducible interpretation of favorable comparisons and detection trade-offs, without converting descriptive results or hypothetical calculations into operational guarantees.}
\maketitle

\section{Introduction}
A useful intrusion-detection comparison must link the false-positive rate (FPR) and attack true-positive rate (TPR) to the same detector, decision rule and evaluation set. A lower false-alarm rate may accompany either improved detection or more missed attacks; overall accuracy alone does not distinguish these outcomes.

This distinction matters operationally for intrusion detection systems (IDS). When attacks are rare, benign misclassifications can dominate flagged decisions: the fraction of flags representing attacks depends on the attack base rate as well as FPR and TPR~\cite{Axelsson}. Evaluation must also address traffic variability, the costs of errors, and the gap between a classified record and an actionable security finding~\cite{Sommer}. Arp et al.~\cite{Arp} further show why evaluation design, baselines and application-appropriate metrics matter when interpreting machine-learning results in security. These considerations motivate examining both rates and their provenance; they do not specify a universal acceptable FPR or justify transporting benchmark rates to deployments.

Reviews of quantum machine learning for intrusion detection (QML-IDS) summarize architectures and metrics~\cite{Kaissar,Sai,Secure}. We ask: \emph{In the eligible studies of the Kaissar corpus, what quantitative evidence can be recovered about false positives and attack detection at the same operating point, and what is the scope of those conclusions given evaluation size and threshold selection?} The 25 entries selected by Kaissar et al. define our closed frame. Its prior quality filter and finite scope permit an auditable reconstruction but preclude treating it as a representative census of QML-IDS.

This original secondary research reconstructs published quantitative evidence rather than providing an exhaustive literature review. Our contribution is the correspondence between published objects, class-specific counts, denominators and provenance, and the comparisons that this correspondence supports. The study adds no detector training, quantum execution or new operating threshold. Standard statistical calculations are used as tools, not claimed as novel methods.

\section{Related work and scope of the contribution}
Kaissar et al.'s Table 2 supplies the sampling frame~\cite{Kaissar}. Sai et al.'s intrusion-detection discussion and Section 5.1/Table 3 already summarize false-positive and detection metrics~\cite{Sai}; reviewing these metrics is therefore not itself new. Their review spans 44 cybersecurity studies, whereas our extraction follows each retained source object to its denominator and evaluation role. The secure-communications review also discusses evaluation maturity across applications~\cite{Secure}. We do not infer source-specific disagreements with those reviews where an identical detector, subset and metric cannot be matched. The principal comparisons below are with the original reports, not a scorecard against review summaries.

Lamberts et al.~\cite{Lamberts} systematically examine industrial IDS evaluation, including metric definitions, confusion-matrix reporting and comparability. Their analysis establishes relevant methodological concerns; our distinct empirical object is the joint reconstruction of rates, denominators and provenance at recoverable operating points in a closed QML-IDS corpus.

CLAIMSTAB-QC~\cite{Claimstab} treats empirical quantum-software comparisons as auditable claims under evaluation contracts. Our contribution is narrower and empirical: recovering joint false-alarm/detection evidence in a defined IDS corpus, including points without a quantum--classical comparator and normalized results that identify only ranges.

Clopper--Pearson limits~\cite{CP} quantify binomial sampling uncertainty under stated assumptions. Neyman--Pearson classification~\cite{Tong} uses independent calibration samples and order statistics to control a population error rate. Assessing a frozen detector and constructing one to meet an error constraint are different operations; neither makes undocumented independence true. Our provenance rule makes that distinction explicit.

DP-QIDS~\cite{DP}, the Meta-Quantum Ensemble (MQE)~\cite{MQE} and Gillani et al.~\cite{Gillani} provide external methodological contrasts. DP-QIDS separates an F1-optimized matrix from fixed-FPR analyses and retains a finite-sample slack term; MQE evaluates low-FPR detection with calibration metrics; Gillani et al. use controlled splits, repeated seeds and paired analyses. These are new detector evaluations, not reconstructions of the selected corpus. Their counts and conclusions remain outside our denominator. More generally, Bowles, Ahmed and Schuld~\cite{Bowles} use classical baselines and component ablations to examine QML benchmarks, illustrating why a model-level comparison alone does not attribute an improvement to its quantum component. We do not transfer their benchmark outcomes to this corpus.

\section{Materials and Methods}
\subsection{Frame, eligibility and units}
Kaissar et al. report a search through January 2026. Their selection starts with 387 records, removes 49 duplicates, screens 338 records, excludes 299, and evaluates 39 full texts. Ten further scope exclusions and four quality exclusions leave 25 articles; their quality criterion is at least four of seven assessment items~\cite{Kaissar}. We inherit that selection, including its scope and publication biases. We do not treat its quality screen as an independent validation of the statistical assumptions required here.

Eligibility requires a published empirical intrusion-detection article with benign-versus-attack relevance and a quantum computational component, executed physically or in simulation. Quantum-inspired classical optimization alone is excluded. Eligibility does not require a confusion matrix, a favorable result, a low-FPR objective, or full-text access. All 25 frame entries retain stable identifiers K01--K25. The two quantum-inspired entries K02 and K05 are excluded~\cite{K02,K05}; eligibility remains technically unresolved for K08 and K09~\cite{K08,K09}. The remaining 21 entries are included~\cite{K01,K03,K04,K06,K07,K10,K11,K12,K13,K14,K15,K16,K17,K18,K19,K20,K21,K22,K23,K24,K25}. K13 and K21 belong to one identified documentary family; the 21 articles therefore represent 20 identified families, not 21 established independent experiments. Unrecognized reuse may remain.

The extraction unit is an article--source version--dataset/subset--detector--operating point. A point means a published decision rule at an evaluated setting, even if a numerical threshold is not disclosed. It is not an independent study. Multiple depths, backends, sizes, training rows and administrative records are retained for traceability without increasing the article denominator. The frozen database contains 466 records, including summary-metric and non-analysable entries; it does not contain 466 independently evaluated operating points.

\subsection{Access, extraction and provenance}
The corpus and extracted source counts were fixed on 24 October 2026. Subsequent revisions refined their interpretation and presentation, including provenance-based interval classification, without changing those counts. This freeze documents the extraction boundary of an exploratory secondary analysis, not a preregistered design. Publisher texts, identified author versions, deposited code and indexed primary text were consulted through legitimate access routes. Substantive methodological material was consulted for 15 of the 21 included articles; this does not imply access to complete methods, all figures or all supplementary artifacts. Six included articles remain without a recovered full methods/results text. SI Table S2 and Section S7 list all 25 entries and document recovered and unrecovered source material. A failed HTTP request is recorded as retrieval failure, not proof of a paywall. Verified restrictions are distinguished from technical failures; an unrecovered figure is not classified as an unreported figure. Chao (K04)~\cite{K04} and Gong (K06)~\cite{K06} remain in the frame without substitutes.

For each result, we record the original claim, dataset name and evaluated subset, class meaning, axes, counts or metrics, aggregation, threshold rule, and the correspondence between these fields. We distinguish named training results from independent tests, nominal split sizes from observed class denominators, and repeated-run averages from a single confusion matrix. Where source objects disagree, we retain separate objects rather than repairing them by assumption. Missing counts are not inferred from rounded accuracy, precision and recall.

An OpenAI Codex assistant supported the extraction design, source retrieval and interpretation, transcription, code development, deterministic calculations, figure generation and manuscript drafting. This use went beyond language editing. A second technical pass checked decisive cells against available source objects and recomputed intervals by independent binomial-tail inversion. Both passes used the assistant and do not constitute independent human extraction. No synthetic source counts, detector predictions or replacement experiments were introduced. The authors confirmed completion of human validation and manuscript approval. This author validation is not documented as independent duplicate extraction; the computational checks address arithmetic and consistency rather than establishing sampling or provenance assumptions. This distinction is consistent with Kapoor and Narayanan~\cite{Kapoor}: reproducing a computation does not by itself establish that the evaluation design supports its scientific interpretation.

\subsection{Rates, resolution and uncertainty}\label{sec:uncertainty}
We code attack as the positive class. For an identified evaluation point, true negatives (TN), false positives (FP), false negatives (FN) and true positives (TP) give
\begin{equation}
 n_0=\TN+\FP,\quad n_1=\TP+\FN,\qquad
 \widehat{\FPR}=\frac{\FP}{n_0},\quad
 \widehat{\TPR}=1-\frac{\FN}{n_1}.
\end{equation}
The empirical resolution of FPR is $1/n_0$: it is a counting increment, not a confidence limit. In a multiclass detector, every non-benign prediction is collapsed to attack; confusion among attack types does not constitute a missed attack. A row-normalized multiclass matrix identifies the benign error rate but generally not the overall TPR without attack-class weights. We report the range over admissible weights, with an allowance for displayed rounding, rather than imposing equal weights.

For identified counts we calculate marginal two-sided 95\% Clopper--Pearson intervals and a one-sided 95\% upper FPR limit. With $x=\FP$ and $n=n_0$, the latter is
\begin{equation}\label{eq:upper}
 U_{.95}(x,n)=B^{-1}_{.95}(x+1,n-x)\quad(x<n),
 \qquad U_{.95}(n,n)=1,
\end{equation}
where $B^{-1}$ is a beta quantile. These calculations condition on independent, identically distributed class-specific observations and a detector/threshold fixed independently of the evaluated sample. Temporal or flow dependence, preprocessing leakage, model selection on test performance, or uncertain dataset provenance can invalidate that interpretation~\cite{Arp,Kapoor}. Their presence cannot be repaired from aggregate matrices. We therefore label the calculations as conditional and do not present them as verified population guarantees. No joint 95\% region or simultaneous coverage across models is asserted.

We use 1\%, 0.1\% and 0.01\% as analyst-defined references separated by orders of magnitude, to show how evidential requirements change across progressively lower false-positive rates. They are analytical scales, not universal operational targets or targets attributed to the source authors. Under the specified fixed-detector binomial model, we ask whether $U_{.95}\le r$ for each reference $r$, conditional on the stated model. A limit above $r$ means that this procedure does not support the bound; it does not prove that the true rate exceeds $r$. We found no explicit numerical low-FPR target in the 15 included articles whose methodology was consulted; targets in inaccessible material remain unknown. Feasibility studies are not judged as failing objectives they never declared.

We apply a provenance rule before displaying intervals. All identifiable rates remain descriptive. Counts described as training results, or accompanied by reported adjustments after observing the test, receive \emph{hypothetical illustrations only}, outside the main interval block. This applies to K11 and K03, respectively. Other eligible count points remain \emph{conditional illustrations with unverified independence}: these are K10 and K17, not a validated independent-test tier. Their inclusion does not establish detector freezing, independent observations or a calibration guarantee. Normalized points without denominators receive identification/rounding ranges only. The provenance categories distinguish documented procedural concerns from unresolved provenance; they do not rank sources by validated independence. The absence of a reported test-based adjustment is not evidence that no such adjustment occurred. Classification may also depend on the extent of source material recovered.

The main conditional analysis comprises six points from K10 and K17. Supplementary Information (SI; \texttt{supplement\_revised.pdf}) examines provenance and joint-matrix completeness as separate sensitivity criteria. The subordinate complete-matrix subset contains five K17 points; this restriction does not challenge the FPR identified by K10's complete benign row. We neither pool rates nor infer field-wide frequencies. Records are not translated into alerts or incidents.

\section{Results}
\subsection{What the corpus permits}
\begin{figure}[p]\centering
\includegraphics[width=\linewidth]{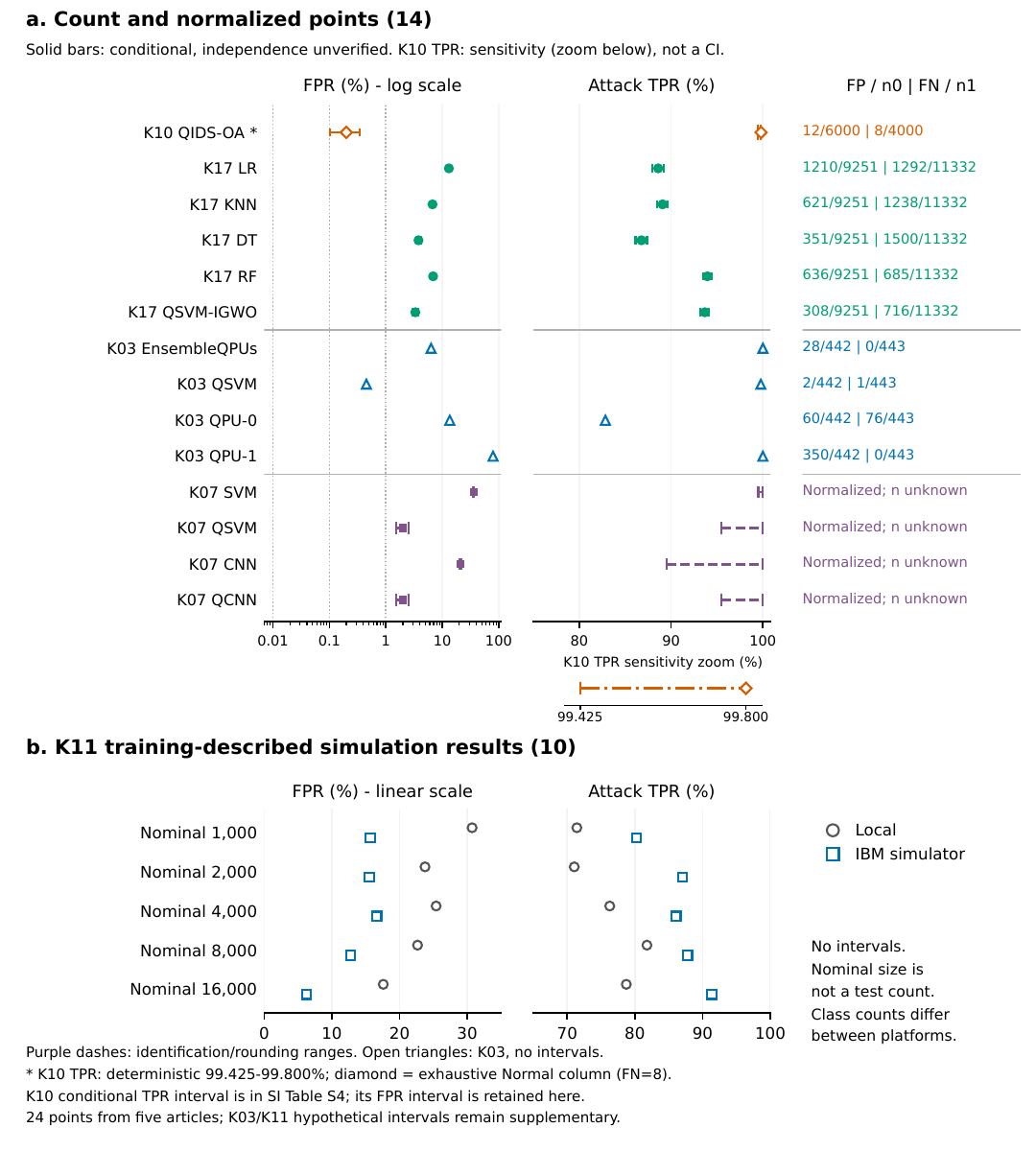}
\caption{All 24 descriptive points from five articles, by provenance. (a) Six K10/K17 FPR intervals and five K17 TPR intervals are marginal 95\% conditional illustrations, with independence unverified. K10 TPR uses an orange dash-dot deterministic sensitivity, 99.425--99.800\%, shown in the zoom inset, with an open diamond at the exhaustive-FN endpoint; conditions are in Section 4.2 and its conditional TPR interval in SI Table S4. K03 test-adjusted points (open triangles) have no intervals. K07 purple dashes show identification/rounding ranges, assuming nearest rounding of proportions to two decimals ($\pm0.5$ percentage points), not sampling uncertainty. Dotted lines denote analytical references. (b) Ten K11 training-described points by nominal size and platform, without intervals or trajectories; matching sizes do not imply shared instances. FPR is logarithmic in a and linear in b. SI retains K03/K11 hypothetical intervals and K11 counts. Source reports: \cite{K03,K07,K10,K11,K17}}\label{fig:main}
\end{figure}

Table~\ref{tab:denom} specifies the denominator at each stage. The 24 operating points from five articles are traceable configurations, not 24 independent replications. Five articles supply the 24 points in Figure~\ref{fig:main}: K03 contributes four, K07 four, K10 one, K11 ten and K17 five. Nineteen points come from complete matrices, one from a complete benign row with incompletely allocated attack rows, and four from normalized matrices. Ten of the nineteen complete matrices are K11 training-described results. Thus completeness of a matrix and validity of independent-test inference are separate properties.

\begin{table}[htbp]\centering\small
\caption{Units and analysis denominators. These rows are nested or overlapping; they must not be added.}\label{tab:denom}
\begin{tabular}{P{13.3cm}r}\toprule
Unit or analysis & Count\\\midrule
Entries in the previously selected Kaissar frame &25\\
Included articles / excluded entries / unresolved eligibility &21 / 2 / 2\\
Identified documentary families among included articles &20\\
Included articles with substantive methodology consulted &15\\
Articles contributing to the main figure &5\\
Main-figure operating points &24\\
Complete matrices / partial matrix / normalized points &19 / 1 / 4\\
Main-figure points described by their source as training &10\\
Conditional illustrations: K10 and K17, independence unverified &6\\
Hypothetical-only count points: K03 and K11 &14\\
Subordinate complete-matrix sensitivity: K17 only &5\\\bottomrule
\end{tabular}
\end{table}

Table~\ref{tab:objects} summarizes the interpretive consequences against the original sources.
\begin{longtable}{P{4.65cm}P{5.35cm}P{6.15cm}}
\caption{From the original source object to its interpretation. These five articles supply the 24 descriptive points; rows are not independent replications. Locations refer to the versions in SI.}\label{tab:objects}\\
\toprule
Published object and location & Reconstructed evidence & Consequence for interpretation\\\midrule\endfirsthead
\toprule
Published object and location & Reconstructed evidence & Consequence for interpretation\\\midrule\endhead

Payares and Martinez-Santos (K03)~\cite{K03}: Section 3, printed pp.~5--7, Figures 6--9.
& Four matrices, each with 442 benign and 443 attack cases; the quantum support vector machine (QSVM) has 2 FP and 1 FN. Section 3 describes adjustments after test inspection.
& Rates remain descriptive; all four intervals are hypothetical and supplementary. The saved hybrid quantum neural network output is a separate, unmatched source object.\\[4pt]
Kalinin and Krundyshev (K07)~\cite{K07}: ``The experimental study'', Figures 5--8.
& Four row-normalized matrices identify the Normal diagonal; attack weights and class counts are unavailable.
& FPR has displayed-rounding ranges; aggregate TPR has weight/rounding ranges. No exact counts or sampling intervals can be assigned.\\[4pt]
Kim and Madhavi (K10)~\cite{K10}: Tables 2--3, Results, PDF p.~9.
& Complete benign row: 12 FP/6000. Declared attack total 4000; explicit attack entries sum to 3985; eight entries predict Normal.
& FPR is identified separately. TPR 99.8\% requires Table 2--3 linkage and exhaustive false negatives. A separately labelled reporting-completeness sensitivity gives 99.425--99.800\%; FPR is unaffected. Binomial illustrations retain unverified independence.\\[4pt]
K\"{u}\c{c}\"{u}kkara et al. (K11)~\cite{K11}: Section 3.3, PDF p.~9; Figures 4--5, pp.~10--11; Table 3, p.~12.
& Ten matrices described as training results; class counts differ between simulation platforms. Matrix-derived recall/precision match reversed table headings.
& Retain all descriptive rates and the labelled matrix axes. Hypothetical intervals stay supplementary; nominal size does not establish an independent test.\\[4pt]
Elsedimy et al. (K17)~\cite{K17}: Figure 7, \emph{Test} matrices, PDF p.~15; Table 4 is a separate object.
& Five Test-labelled points, each with 9251 benign and 11332 attack cases; QSVM has lower FPR than all four classical comparators.
& TPR improves against three comparators and decreases against random forest. Table 4 cross-validation means and Train matrices are not substituted. Equal denominators establish neither shared instances nor paired inference.\\\bottomrule
\end{longtable}

The other ten consulted articles remain informative without supporting a primary joint point. Their evidence includes unidentified class counts, F1-only summaries, normalized aggregate metrics, unresolved class coding or axis orientation, and unavailable matrix images. K15~\cite{K15} supplies four count matrices whose benign/attack mapping remains conditional; K16~\cite{K16} supplies normalized matrices with unresolved orientation. These are explicit supplementary sensitivities. K23~\cite{K23} reports figures whose cells were not recovered. None of these situations licenses fabricated counts, and the six included articles without full text are not counted as non-reporting studies.

\subsection{Low observed FPR does not identify the same inferential claim}
K03~\cite{K03} contains four published matrices on 442 benign and 443 attack examples. The quantum support vector machine (QSVM) matrix identifies two false positives and one false negative: FPR 0.452\%, TPR 99.774\%. The ensemble identifies 28 false positives and no false negatives: 6.335\% and 100\%. Thus the QSVM's lower false-alarm rate accompanies one additional missed attack, rather than an improvement in every measure. If these were independent observations of a fixed detector, the one-sided upper FPR limit would be 1.417\%, despite the observed rate below 1\%. Because Section 3 describes test-based adjustments, this calculation is hypothetical and is retained only in SI. The simulator submodels named QPU-0 and QPU-1 (QPU: quantum processing unit) have FPR 13.575\% and 79.186\%, respectively; those names do not establish physical quantum execution. The saved hybrid quantum neural network (H-QNN) notebook output has zero false positives among 442 benign examples and would give $U_{.95}=0.675\%$ under that hypothetical model. Its identity with the published figure has not been established, so it remains supplementary and does not enter the primary published-point calculation. Preprocessing and adjustments described after test inspection also prevent unqualified independent-test inference.

K10~\cite{K10} reports a complete Normal row: 5988 correct Normal decisions and 12 predictions assigned to attack classes, among 6000 benign nodes/instances. This identifies FPR 0.200\% and a conditional upper limit of 0.324\%. The four attack rows do not fully allocate their declared 4000 cases: their explicit entries total 3985. The Normal column contains eight attack-to-Normal errors. Consequently, TPR 99.800\% additionally requires linking the 4000 attacks in Table 2 to Table 3 and treating that column as exhaustive. We do not fill the fifteen unallocated cases. Assuming that Tables 2 and 3 refer to the same 4000 attack cases and that the 3985 explicitly allocated outcomes are correct, we consider a sensitivity scenario in which the 15 remaining cases may have any binary prediction. Writing $m\in\{0,\ldots,15\}$ for additional false negatives gives $\TPR(m)=1-(8+m)/4000$, or 99.425--99.800\%. The Normal column is numeric in all four attack rows; the dashes occur in other attack-class columns. Allowing $m>0$ therefore relaxes the exhaustiveness of the Normal column rather than merely filling dashed attack-to-attack cells. This deterministic range is not a confidence interval and does not assert that additional FN were observed. If the numeric Normal column is exhaustive, the conditional point remains 99.800\%; its binomial interval in SI Table S4 does not incorporate the documentary uncertainty. Without the assumed table correspondence, or with other unbounded reporting errors, the attack-side range is not identified. These attack-side conditions do not alter the complete benign row or remove its FPR from the main analysis. The nominal 100000-record split elsewhere in the article is not substituted for these matrix denominators.

Under the fixed-detector binomial model, the six-point block contains K10 and the five K17 Test-labelled matrices; independence remains unverified for both studies. K10 is the only point in this block with $U_{.95}<1\%$. None has an upper limit at or below 0.1\% or 0.01\%. The subordinate complete-matrix restriction leaves five K17 points and no sub-1\% limit; it does not negate K10's recoverable FPR.

The benign denominators in this block are 6000 and 9251. Even with zero false positives, its maximum denominator, 9251, cannot support the 0.01\% analytical reference using Equation~\ref{eq:upper}: at least 29956 benign observations would be needed. The corresponding zero-error requirements for 1\% and 0.1\% are 299 and 2995. This is a limitation of the specified upper-limit procedure at these sample sizes, not evidence that a true rate exceeds a reference or that a source failed its own objective.

\subsection{Favorable comparisons and detection trade-offs}
\begin{figure}[htbp]\centering
\includegraphics[width=.84\textwidth]{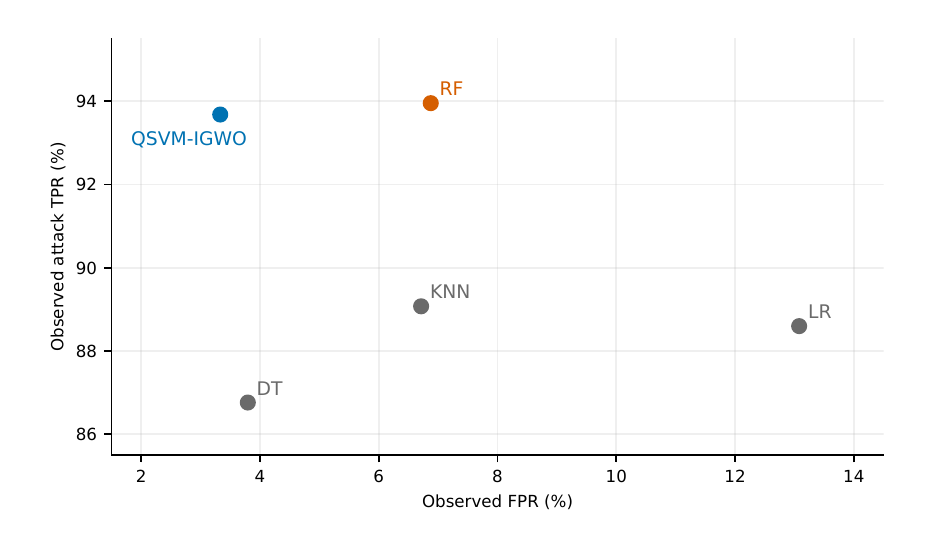}
\caption{All five Test-labelled K17 operating points. Lower FPR and higher TPR are favorable. QSVM improves both displayed rates relative to LR, KNN and DT, but trades lower FPR for slightly lower TPR relative to RF. These are separate detector points, not samples of one receiver operating characteristic (ROC) curve. No paired uncertainty or verified identity of evaluated instances is implied. Source report: \cite{K17}}\label{fig:k17}
\end{figure}

K17~\cite{K17} provides five matrices explicitly labelled Test, each with 9251 benign and 11332 attack cases. The QSVM with improved grey wolf optimization (QSVM-IGWO) point has 308 false positives and 716 false negatives; the random forest (RF) point has 636 and 685. The reconstructed change is therefore $-3.546$ percentage points in FPR and $-0.274$ in TPR (Figure~\ref{fig:k17}). In descriptive terms, the quantum model reduces false alarms while missing slightly more attacks. The relative FPR decrease is approximately 51.6\%, but this is not an estimate of operational alert reduction.

QSVM also has lower FPR than logistic regression (LR), k-nearest neighbours (KNN) and decision tree (DT). Against these three it has higher TPR: improvements of 5.083, 4.606 and 6.918 percentage points, respectively. The favorable QSVM--classical comparisons therefore survive reconstruction, with RF revealing a distinct trade-off. 

Matching denominators do not establish that the same instances were evaluated, and predictions are unavailable for a paired analysis. We report no McNemar test, paired confidence interval or statistically established dominance. The Figure 7 counts imply a 75/25 Train/Test allocation, whereas the article describes an 80/20 split and separately reports ten-fold means. Its narrative counts 33942 and 58 belong to the training matrix, with 58 denoting false negatives. We preserve the Test-labelled matrices, the training matrices and the cross-validation means as separate source objects. We retain BoT-IoT as the dataset name declared by the source; correspondence between the matrices and records from an identified version of that dataset has not been established.

K07~\cite{K07} supports a different kind of comparison. Its row-normalized matrices give approximate FPRs of 36\% for a support vector machine (SVM), 2\% for QSVM, 21\% for a convolutional neural network (CNN) and 2\% for its quantum counterpart (QCNN). This confirms improved recognition of Normal traffic at the reported settings. Overall attack TPR is not uniquely recoverable: without attack weights, displayed-row bounds are 96--100\% for QSVM/QCNN and 90--100\% for CNN, whereas SVM has no displayed attack-to-Normal errors. Figure~\ref{fig:main} expands these TPR ranges for rounding and displays the FPR rounding ranges: 35.5--36.5\%, 1.5--2.5\%, 20.5--21.5\% and 1.5--2.5\%, respectively, under rounding to nearest. Misclassification of one attack type as another is not counted as a binary missed attack. No binomial interval or exact pooled TPR is assigned to these normalized results.

\subsection{Training descriptions and size comparisons}
K11~\cite{K11} supplies ten complete matrices across five nominal sizes and two simulation platforms. Section 3.3 describes the local and quantum-simulator matrices as training results. Each matrix happens to total 30\% of its nominal size, but that arithmetic does not establish a held-out evaluation. Accordingly, all ten points are descriptive in the main figure; SI Table S8 provides their counts and hypothetical calculations.

From the smallest to largest nominal size, the local quantum neural network (QNN)'s displayed FPR decreases from 30.719\% to 17.599\% and TPR increases from 71.429\% to 78.734\%. The corresponding IBM-simulator values move from 15.686\% to 6.262\% and from 80.272\% to 91.375\%. Intermediate settings are retained, and this is not a causal estimate of a sample-size effect. Class counts differ between platforms at several nominal sizes, so neither shared instances nor a paired platform comparison is established. The source's Table 3 precision and recall columns numerically match matrix-derived recall and precision, respectively. We use the labelled matrix axes and disclose the discrepancy. These points compare simulation settings of a QNN, not quantum and classical detectors.

\section{Discussion}
\subsection{What the reconstruction adds}
Table~\ref{tab:objects} distinguishes three questions that a single reporting-quality label would conflate: what the source matrix identifies, what comparison follows, and what sampling interpretation its provenance permits. Favorable comparisons survive reconstruction, but a lower false-alarm rate need not improve both outcomes. The division between rate identification and inferential provenance also explains why a complete benign row can remain informative while complete training or test-adjusted matrices provide only descriptive evaluation evidence.

The uniform rule for K03 and K11 prevents a test label from overriding contrary procedural evidence. Conversely, retaining K10 and K17 in a conditional illustration block does not certify their independence.

The evidence does not establish quantum advantage at a controlled FPR, a common deployment distribution, or a matched resource budget. The observed comparisons concern particular detectors and datasets. Different thresholds could change their trade-offs, but a confusion matrix at one threshold cannot identify those alternatives. Likewise, lack of a dedicated FPR-calibration method does not automatically invalidate a well-designed independent test. The relevant question is whether independence, selection and sampling assumptions are supported for the reported point.

\subsection{Implications for evaluation practice}
A reusable report should link one matrix or score file to a named evaluation subset, class definitions, the selected detector and threshold, and the data used for selection. Publishing class counts and provenance alongside FPR and TPR would often suffice to make a comparison interpretable; a new statistical method is not required. For controlled low-FPR claims, an independent benign calibration sample and a documented decision rule can provide a stronger contract~\cite{Tong,DP}. Calibration error metrics alone are not such a guarantee, and a distributional bound must retain its finite-sample slack.

Dependence matters particularly for traffic data. Randomly splitting related flows, packets, hosts or sessions may yield many records without the same number of independent observations. Aggregate matrices cannot identify the resulting effective sample size. The counts in this study retain the source's unit; nodes, traffic records, alerts and incidents are not equated. A hypothetical volume calculation would require a specified benign exposure and stable transported conditional error rate, and would still predict misclassified units rather than distinct security incidents. We leave that translation outside the principal analysis.

\subsection{Limitations}
The frame is deliberately narrow and selected previously for quality. It is not a probability sample of QML-IDS, and the recoverability fractions cannot be generalized to the field. Six included articles lack recovered methodological text, two frame entries have unresolved eligibility, and one consulted article has unrecovered matrix images. These access limitations may be informative about neither reporting quality nor performance. The five articles yielding the main figure are a particularly selected subset of the frame.

The quantitative core is further limited: of 24 points, ten are training-described, four accompany reported test-based adjustments, and four are normalized rather than raw counts. The main conditional illustrations concern six points from two articles; their independence remains unverified. Multiple points within an article are dependent evidence, and even documentary families are not guaranteed independent experiments. No pooled efficacy estimate or cross-study significance test is justified.

Thresholds and evaluation provenance are not uniformly established. Conditional intervals illustrate what specified counts would imply under a fixed-detector binomial model; none is labelled a verified deployment or calibration guarantee. Finally, the analysis is exploratory secondary research: the frozen extraction and explicit sensitivities support reproducibility but do not turn the earlier feasibility work into a prospective confirmatory study.

\section{Conclusions}
Recovering source objects and denominators makes favorable results more specific. The QSVM comparison in K17 combines lower FPR against all four classical detectors with a small detection loss against random forest and detection gains against the others. K10's benign row identifies its FPR independently of the extra assumptions needed for attack detection. Normalized matrices in K07 support ranges rather than an invented aggregate TPR.

These findings concern 24 descriptive points from five articles within a closed corpus of 21 included articles and 20 identified documentary families. All rates are retained, while the same provenance rule places K03 and K11 intervals in a hypothetical supplement. Six K10/K17 points remain conditional illustrations, not verified independent evaluations. Their sample sizes cannot support the 0.01\% reference with the specified upper limit even in a zero-error case. The resulting contribution is a reproducible account of which published comparisons and detection trade-offs are identifiable, and which additional interpretations require assumptions that the available evidence does not establish.

\section*{Supplementary Information}
The accompanying file \texttt{supplement\_revised.pdf} (Supplementary Information, SI) contains the extraction rules, all 25 source dossiers, conditional calculations, provenance and matrix-completeness sensitivities, and the correspondence to the reproducible artifact. It is separate from the manuscript and forms part of the submission.

\section*{Declarations}
\subsection*{Author Contributions}
Investigation, R.F.-B.; writing---original draft, R.F.-B.; writing---review and editing, I.P.-L., A.G.-S. and P.G.B.; supervision, I.P.-L., A.G.-S. and P.G.B. All authors have read and approved the manuscript.

\subsection*{Funding}
This research received no specific external funding.

\subsection*{Competing Interests}
The authors declare that they have no competing interests.

\subsection*{Data Availability}
The derived records and source/version locators supporting this study are archived in version 2.2.0 of the reproducibility artifact at \url{https://doi.org/10.5281/zenodo.23642804}. Derived data and documentation are licensed under CC BY 4.0. Original reports are identified by DOI and source locations; their copies are retained privately and are not redistributed. The additional K10 sensitivity calculation and revised Figure 1 sources are provided as Supplementary Code (\texttt{revision\_materials.zip}) with this manuscript.

\subsection*{Code Availability}
The analysis code, portable checks and figure/table sources are available in the same version 2.2.0 archive~\cite{Artifact} and at \url{https://github.com/roberto-fernandez-barrios/qml-ids-operating-points/releases/tag/v2.2.0}; the repository is \url{https://github.com/roberto-fernandez-barrios/qml-ids-operating-points}. The baseline entry point is \path{scripts/reproduce.py}. That archive reproduces the baseline numerical results and figure content; it does not generate the revised Figure 1. Supplementary Code (\texttt{revision\_materials.zip}) supplies the additional K10 sensitivity calculation, its checks and the revised Figure 1 source; its entry point is \path{reproduce_revision.py}. The original corpus and extracted counts are unchanged. Code is MIT-licensed.

\subsection*{Ethics and Consent}
Ethics approval and consent to participate or publish are not applicable to this secondary analysis of published aggregate results; no new human-participant or animal study was conducted.

Consent to Publish declaration: not applicable

\subsection*{Acknowledgements and Use of Artificial Intelligence}
OpenAI Codex was used to assist with source review and data extraction, analysis code and checks, figure generation, and manuscript drafting and editing. Claude Code (Anthropic), running Claude Fable 5.1, was used for critical review and consistency checking. The authors reviewed and approved the final content and take full responsibility for the manuscript.

\subsection*{Author Identifiers}
Roberto Fern\'andez-Barrios: \url{https://orcid.org/0009-0003-5312-2634}; Iker Pastor-L\'opez: \url{https://orcid.org/0000-0002-3068-6248}; Asier Gonz\'alez-Santocildes: \url{https://orcid.org/0009-0002-8888-8560}; Pablo Garc\'ia Bringas: \url{https://orcid.org/0000-0003-3594-9534}.

\end{document}